\documentclass[prd,aps,twocolumn,a4paper,floatfix,showpacs,nofootinbib]{revtex4-1}
\usepackage[utf8]{inputenc}  
\usepackage[T1]{fontenc}
\usepackage{graphicx,psfrag}
\graphicspath{{figures/}}
\usepackage{mathrsfs}
\usepackage{amsmath,amsfonts,amssymb}
\usepackage{multirow,enumerate}
\usepackage{comment,hyperref}
\usepackage{color}
\usepackage{acronym}
\usepackage{xspace}
\usepackage[normalem]{ulem}
\usepackage{mathtools}
\usepackage{subfigure}
\usepackage{makecell}
\usepackage{appendix}
\usepackage{diagbox}
\usepackage{lineno}
\usepackage{textcase}
\usepackage{hyperref}

\newacro{BH}{black hole}
\newacro{NS}{neutron star}
\newacro{PN}{Post-Newtonian}
\newacro{BBH}{binary black hole}
\newacro{BNS}{binary neutron star}
\newacro{EOB}{effective-one-body}
\newacro{NR}{numerical relativity}
\newacro{GW}{gravitational wave}
\newacro{EOS}{equation-of-state}

\newcommand{\be}{\begin{equation}}
\newcommand{\ee}{\end{equation}}
\newcommand{\bea}{\begin{eqnarray}}
\newcommand{\eea}{\end{eqnarray}}
\newcommand{\bel}{\begin{align}}
\newcommand{\eel}{\end{align}}

\def\i{{\rm i}}

\def\GMc2{{\rm G M_{\odot} c^{-2}}}

\def\SEOBNRv4T{\texttt{SEOBNRv4T}\xspace}

\def\DGW{D_L^\textrm{GW}}
\def\DEM{D_L^\textrm{EM}}
\def\betaMGR{\vec{\theta}_\textrm{MGR}}

\usepackage{color}
\definecolor{cyan}{rgb}{0,0.9,0.9}
\definecolor{orange}{rgb}{0.9,0.5,0}
\definecolor{magenta}{rgb}{1,0,1}
\definecolor{purple}{rgb}{0.8,0.4,0.8}
\definecolor{gray}{rgb}{0.5,0.5,0.5}
\definecolor{mygreen}{rgb}{0.1,0.8,0.1}
\definecolor{darkblue}{rgb}{0.0,0.0,0.6}

\definecolor{mangotango}{rgb}{1.0, 0.51, 0.26}

\begin{document}

\title{How well can modified gravitational wave propagation be constrained with strong lensing?}

\author{Harsh Narola$^{1,2}$}
\author{Justin Janquart$^{1,2}$}
\author{Leïla Haegel$^{3}$}
\author{K. Haris$^{1,2,4}$}
\author{Otto~A.~Hannuksela$^{5}$}
\author{Chris Van Den Broeck$^{1,2}$}

\affiliation{${}^1$Institute for Gravitational and Subatomic Physics (GRASP),
	Utrecht University, Princetonplein 1, 3584 CC Utrecht, The Netherlands}
\affiliation{${}^2$Nikhef -- National Institute for Subatomic Physics,
	Science Park 105, 1098 XG Amsterdam, The Netherlands}

\affiliation{${}^3$Universit\'e Paris Cit\'e, CNRS, Astroparticule et Cosmologie, F-75013 Paris, France}

\affiliation{${}^4$Department of Physics, National Institute of Technology, Kozhikode, Kerala 673601, India}

\affiliation{${}^5$Department of Physics, The Chinese University of Hong Kong, Shatin, New Territories, Hong Kong}

\date{\today}

\begin{abstract}
Strong gravitational lensing produces multiple images of a gravitational wave (GW) signal, which can be 
observed by detectors as time-separated copies of the same event. It has been shown that under 
favourable circumstances, by combining information from a quadruply lensed GW with electromagnetic 
observations of lensed galaxies, it is possible to identify the host galaxy of a binary black hole 
coalescence. Comparing the luminosity distance obtained through electromagnetic means with 
the effective luminosity distance inferred from the lensed GW signal would then enable us to constrain alternative 
theories of gravity that allow for modified GW propagation. Here we analyze models including large extra 
spatial dimensions, a running Planck mass, and a model that captures propagation effects occurring 
in a variety of alternative theories to general relativity. 
We consider a plausible population of lenses and binary black holes and use Bayesian inference 
on simulated GW signals as seen in current detectors at design sensitivity,  
to arrive at a realistic assessment of the bounds that could be placed. We find that, due 
to the fact that the sources of lensed events will typically be at much larger redshifts,  
this method can improve over bounds from GW170817 and its electromagnetic counterpart by a factor of 
$\sim 5$ to $\mathcal{O}(10^2)$, depending on the alternative gravity model.   



\end{abstract}

\maketitle

\section{I\lowercase{ntroduction}}
\label{sec:intro}

Since the first direct detection of gravitational waves (GWs) in 2015, the field of GW physics has been developing rapidly~\cite{LIGOScientific:2016aoc}. 
The network of two Advanced LIGO detectors~\cite{LIGOScientific:2014pky} and one Advanced 
Virgo detector~\cite{VIRGO:2014yos} has observed around 90 GW signals to date~\cite{LIGOScientific:2021djp}. These observations have opened up several 
previously unexplored research directions.
For example, they have led to enhanced tests of general relativity (GR) by providing access to the genuinely strong-field dynamics of spacetime~\cite{LIGOScientific:2021sio}, provided a new method for probing the expansion of the Universe~\cite{LIGOScientific:2021aug}, and 
contributed to a better understanding of the formation channels of the binaries and other astrophysical compact objects~\cite{LIGOScientific:2021psn}.
As the interferometers' sensitivities improve and new detectors such as KAGRA~\cite{Somiya:2011np, Aso:2013eba, KAGRA:2018plz, KAGRA:2020tym} and
LIGO-India~\cite{Unnikrishnan:2013qwa} join the network, even more events will be observed.

The detector upgrades could enable the detection of new phenomena, such as the gravitational lensing of 
GWs~\cite{Takahashi2003, Takahashi2005, Oguri2022AmplitudeLensing}. 
The latter occurs when GWs experience deflection due to a massive object, known as the lens, 
in their path.
Recent rate estimates suggest that GW lensing can become detectable at the rate of $~\mathcal{O}(1)$ per year 
with current deetectors at design sensitivity~\cite{Wierda:2021upe, Ng:2017yiu, Xu:2021bfn, Oguri:2018muv, Mukherjee:2021qam, Smith2022DiscoveringObservatory, Li:2018prc}.
If the Schwarzschild radius of the lens is much larger than the GW wavelength (i.e.~when 
the geometric optics limit applies), it can split the 
observed GW signal into multiple copies, also referred to as the lensed images.
This phenomenon is called the strong lensing of gravitational waves.
The images reach the detector as repeated and time-separated copies of the GW signals that only differ in their amplitudes (due to being magnified/demagnified by the lens), 
overall phases (due to image inversion along one or two principal axes), 
and arrival times (as the images travel along trajectories of different length)~\cite{Dai:2017huk, Ezquiaga:2020gdt}.
By contrast, if the size of the lens is comparable to the wavelength of the GW (referred to as the wave optics limit), 
the GW can undergo frequency-dependent modulation~\cite{Lai2018DiscoveringLensing,Christian2018DetectingObservatories,Singh2018ClassifyingLearning,Dai2018DetectingWaves,Kim2020IdentificationLearning,Cheung2020Stellar-massWaves,Yeung2021MicrolensingMacroimages,Bulashenko2021LensingPattern,Seo2021StrongMicrolensing,Tambalo2022LensingLenses, Wright:2021cbn}.

Gravitational lensing has several interesting applications in fundamental physics, cosmology, and astrophysics 
(for example see~\cite{Deguchi1986DiffractionMass,  Baker:2016reh, Collett2016TestingLensing, Liao2017PrecisionSignals, Fan2017SpeedSignals, Liao2018AnomaliesSubstructures, Diego2019ConstrainingFrequencies,  Fan:2016swi, Lai:2018rto, Cremonese2021BreakingEvents, Mukherjee2019Multi-messengerWaves,Mukherjee2019ProbingSurveys, Oguri2020ProbingWaves, Goyal:2020bkm, Birrer:2022chj, Basak2022ConstraintsMicrolensing, Ezquiaga:2022nak, Wempe:2022zlk, Hannuksela:2020xor, Tambalo:2022wlm, Vegetti:2023mgp, Ezquiaga:2020dao, Goyal:2023uvm}). 
Our work focuses on strong lensing and its ability to test GW propagation beyond GR. 
In particular, it enables tests of theories and models with modified GW propagation. Here 
we will focus on three different (classes of) models: one which has large extra spatial dimensions \cite{Deffayet:2007kf}; one 
where anomalous propagation arises from a time-varying Planck mass \cite{Lagos:2019kds};
and another one proposed in \cite{Belgacem:2018lbp} which captures propagation effects in a number of 
alternative theories of gravity \cite{LISACosmologyWorkingGroup:2019mwx}, 
and which here we will refer to as $\Xi$-parameterization. Recent studies have
already demonstrated that the latter can be tested using strongly lensed events ~\cite{Finke:2021znb}.
Here we provide a comprehensive assessment of the constraints that can be placed on all of the above 
mentioned models, assuming realistic distributions for the parameters characterizing the 
lenses and the binary black holes, for second-generation GW detectors at design sensitivity.

A strongly lensed GW source will have an improved sky localization compared to a non-lensed source, as we can observe the former multiple times with different detector orientations~\cite{Janquart:2021qov, Janquart:2023osz, Lo:2021nae, Hannuksela2020LocalizingLensing, Seto2004StrongLISA}. 
Especially with four detectable images\footnote{30\% of strongly lensed events are predicted to be quadruplets~\cite{Li:2018prc}.}, we may be able to localize the source within $\mathcal O(1) $ square degrees~\cite{Hannuksela:2020xor, Wempe:2022zlk}. 
When the GW source is lensed, we can expect that the electromagnetic (EM) radiation coming from its host 
galaxy is also lensed, as is widely assumed in cosmography 
studies~\cite{Chen:2017rfc, LIGOScientific:2018gmd, Gray:2019ksv, LIGOScientific:2019zcs, DES:2019ccw}. 
A joint GW+EM analysis can help locate the source's host galaxy once its location is narrowed down to a 
few square degrees using only GW data.
In this step, one reconstructs all the lenses in the region provided by the GW data to find which lens 
could best produce a GW quadruplet with 
properties similar to the ones observed; the galaxy that is undergoing lensing by this particular lens is then likely to be the host galaxy of the GW event.
This method was proposed and studied in~\cite{Hannuksela:2020xor, Wempe:2022zlk}.
Once the host galaxy is known, a dedicated spectroscopic or photometric follow-up can lead us to the redshift of the source.
By combining the source's redshift with a cosmological model, we can estimate the source's luminosity distance in a way 
that is unaffected by the anomalous GW propagation~\cite{Hogg:1999ad}.
In addition, we can have another, independent measurement of the source's luminosity distance from the GW data, which could be affected 
by anomalous propagation; by comparing the two distances the anomaly can be discovered or bounded.

Let us denote by $\DEM$ the luminosity distance derived from the EM redshift measurements and a cosmological model, which we will refer to as the EM luminosity distance.
Similarly, let us write $\DGW$ for the luminosity distance measured from the GW data 
when assuming an amplitude fall-off proportional to $1/\DGW$, and call it the GW luminosity distance. In GR, $\DGW$ and $\DEM$ 
coincide, but in alternative theories of gravity there can be a non-trivial relationship between the two. This relationship will be sensitive both to
parameters associated with the deviation from GR,  and to the cosmological parameters. For definiteness, in this work we will 
generally consider a spatially flat Friedmann-Lemaitre-Robertson-Walker (FLRW) Universe 
with cosmological 
constant and negligible radiation density, in which case the cosmological parameters are the Hubble constant $H_0$, 
and the densities of matter and dark energy relative to the critical density, respectively denoted by $\Omega_m$  and $\Omega_\Lambda$. 
For the purposes of this study, we will fix $\Omega_m$ and $\Omega_\Lambda$ to their values
from Planck 2018~\cite{Planck:2018vyg}, whereas $H_0$ will be left free. 
Note that in the relationship between $\DGW$ and $\DEM$ there will be a degeneracy between the deviation parameters and $H_0$~\cite{Lagos:2019kds}. 
Thus, bounds on the deviation parameters
will be determined by the prior information we have from previous measurements on $H_0$, together with 
the measurement uncertainty on $\DGW$. 
For $H_0$, we could in principle choose a fairly narrow prior range informed by the Planck~\cite{Planck:2006aa}, SHoES~\cite{Riess:2021jrx}, or other previous 
measurements~\cite{Wong:2019kwg}. 
However, in our setting, information about $H_0$ can be obtained from the difference in times of arrival of the GW images, together with lens reconstruction through 
electromagnetic means, as explained in detail in~\cite{Hannuksela:2020xor,Wempe2022AObservations}. 
Since the latter will typically lead to wider ranges for $H_0$ compared to the previous $H_0$ measurements, our predictions for the bounds one can obtain 
on the deviation parameters will be on the conservative side. 

Studying modified propagation theories in the context of strongly lensed and localized GW events, especially from 
binary black hole (BBH) coalescences, is attractive, 
because such events can be detected at a higher redshift compared to binary neutron star (BNS) events. 
In the past, modified propagation theories have been tested using 
GW170817 \cite{Mastrogiovanni:2020mvm, Lagos:2019kds, LIGOScientific:2018dkp, Pardo:2018ipy}, 
a signal from a BNS inspiral with an identifiable EM counterpart 
\cite{LIGOScientific:2017vwq, LIGOScientific:2017ync}. However, by cosmological standards, the GW170817 signal 
travelled only a small distance before it reached the detectors, and 
in modified propagation theories, the imprint of the deviation tends to accumulate with distance. 
Other methods have been proposed that exploit the population properties of BBH coalescences observed with GWs 
\cite{Ezquiaga:2021ayr,Leyde:2022orh,MaganaHernandez:2021zyc}; since BBHs can be detected out to larger 
distances, this enables considerably improved bounds over the ones from GW170817. 
 Due to magnification, GWs from \emph{lensed} BBH events can potentially be seen out to redshifts 
 $z \sim 6$~\cite{Wierda:2021upe}, so that more stringent constraints can be expected also from this methodology. 
The aim of this paper is to quantify the gain from GW lensing for 
the different anomalous propagation scenarios considered.

The rest of the paper is structured as follows. In Sec.~\ref{sec:gw-lensing}, we recall the basics of GW lensing. Modified propagation theories are discussed in Sec.~\ref{sec:mpt}, and 
our method for constraining anomalous propagation through lensing is described in Sec.~\ref{sec:methods}.  
Results and comparisons with measurements on GW170817 and other techniques are presented in 
Sec.~\ref{sec:results}. Finally, Sec.~\ref{sec:conclusions} provides conclusions and future directions. 
We work in the geometric unit system so that the speed of light and the gravitational constant are set 
to unity.

\section{G\lowercase{ravitational-wave lensing and distance measurements}}
\label{sec:gw-lensing}

To understand how strongly lensed GWs can be applied to test theories with modified GW propagation, here we briefly summarize the important elements of strong lensing 
(for a detailed overview of GW lensing, see~\cite{Takahashi2003}, and
to understand the localization aspects, see~\cite{Hannuksela:2020xor, Wempe:2022zlk}). 
We will assume that the GW is originating from a BBH coalescence and that it is strongly lensed by a galaxy, one of the most common configurations according to forecasts~\cite{Wierda:2021upe,Smith2022DiscoveringObservatory}.
In such a scenario, the geometric optics limit applies and multiple images of the GWs are produced.

Strong lensing introduces a magnification $\mu_i$, a time delay $t_i^d$, and an overall complex phase shift $\pi n_i$, called the Morse phase, to each image. They modify the waveform as
\begin{equation}
	h_L^{i}(f;\vec{\theta}, \mu_i, t_i^d, n_i) = |\mu_i|^{1/2} e^{i 2 \pi f t_i^d - i \pi n_i} h(f;\vec{\theta})\,,
\end{equation}
where $h_L^i$ is the waveform associated with the $i^{th}$ lensed image, $h(f;{\vec{\theta}})$ is the waveform in the absence of lensing, $f$ is the frequency, and $\vec{\theta}$ are the source parameters of the binary.
The magnifications, time delays, and Morse phases can be calculated by solving the lens equation 
if we have information about the source position and lens properties.

If there is no complementary EM information available, it is not possible to disentangle the 
luminosity distance and magnifications just using GW data, 
as both only appear in the amplitudes of the images, and different images have different magnifications that 
are \emph{a priori} unknown. 
For a given image we usually absorb the magnification into an \emph{effective} GW luminosity distance 
$D_L^{\rm eff, i} = D_L^{\rm GW}/\sqrt{|\mu_i|}$.
However, when EM information is at hand the magnifications can, in in principle, be separately measured through 
lens reconstruction~\cite{Birrer:2018xgm}, at least for quadruply lensed events. 

Suppose we have detected multiple images of a strongly lensed GW with a network of detectors. 
In this scenario, due to Earth's rotation in between the arrival of the different images, the same event is 
observed multiple times with different detector network orientations, allowing for high-accuracy sky 
localization~\cite{Janquart:2021qov, Lo:2021nae}. Since at least a portion of the host galaxy of the BBH coalescence 
must itself be lensed, one can then consider the strongly lensed galaxies in the sky error box
obtained from the GW measurements~\cite{Hannuksela:2020xor}. For each of these one can use the lensed EM image 
fluxes to reconstruct the profile of the lens. By requiring consistency with the GW relative time delays, relative 
magnifications, and Morse phases, one can filter out incorrect lenses and in principle pinpoint the correct lens and 
host galaxy. From spectroscopic or photometric measurements, the redshift of the host galaxy can be obtained. 
Moreover, for quadruply lensed events, the relative time delays of the GW images together with 
the EM reconstruction of the now identified lens, enable measurement of the \emph{absolute} magnifications 
$\mu_i$ \cite{Hannuksela:2020xor}. Combined with GW measurements of $D_L^{\rm eff, i}$ for the different images, 
this leads to a measurement of $\DGW$. 
 
The details about the EM follow-up and its feasibility are documented 
in~\citet{Hannuksela:2020xor} and~\citet{Wempe:2022zlk}. 
Here we consider a scenario where a quadruply lensed GW has already been detected and 
the host galaxy and lens have been identified and characterized, from which we obtain a 
measurement of $\DGW$ as well as a source redshift. 
By combining the redshift measurement with a cosmology we obtain $\DEM$.
The two distance measurements, $\DGW$ and $\DEM$, are then used to test the modified propagation theories. 

In this work, for definiteness we will assume a flat FLRW universe, in which case one has
\begin{equation}
	\label{eqn:dl-em}
	\DEM =\frac{(1+z_s)}{H_0}\int_0^{z_s} \frac{dz'}{E(z')} \, ,
\end{equation}
where $z_s$ is the redshift of the host galaxy, and $E(z) \equiv \sqrt{\Omega_m (1+z)^3+\Omega_\Lambda}$; 
here $\Omega_m$ and $\Omega_\Lambda$ are the matter and dark energy density parameters, and 
$H_0$ is the Hubble constant.

To simulate strongly lensed GWs, we follow~\citet{Wierda:2021upe} and sample BBHs from a \textsc{PowerLaw+Peak} distribution~\cite{KAGRA:2021duu}, strongly lensed by a population of galaxy lenses following the SDSS galaxy catalogue~\cite{SDSS:2017yll}. 
Our network of detectors consists of the two Advanced LIGO interferometers~\cite{LIGOScientific:2014pky}, 
Advanced Virgo~\cite{VIRGO:2014yos}, KAGRA~\cite{KAGRA:2020tym} and LIGO-India~\cite{Unnikrishnan:2013qwa},  
all at design sensitivity. The noise curves of all detectors are implemented using the \texttt{bilby.gw.detector} module of the \textsc{Bilby} (version 1.2.1) software package~\cite{Ashton2018Bilby:Astronomy}. 
The events with network signal-to-ratio (SNR) above 8 are considered detected~\cite{KAGRA:2013rdx}. 
We then estimate the parameters of the simulated events using \textsc{Golum}~\citep{Janquart:2021qov, Janquart:2023osz}, which gives us the effective/measured luminosity distances of each image $D_L^{\rm eff,i}$ as well as the arrival times.
Typically, lens modelling errors and substructure effects will lead to an error budget for the magnification estimates, with $\sim 10\, \%$ standard deviation being a reasonable estimate~\cite{Hannuksela:2020xor, Wempe:2022zlk}. 
Thus, for each GW measurement, we assume that the magnification posterior derived from the EM 
band is given by $p(\mu_i|\vec{d}_{\rm EM}) = \mathcal N(\mu_i|\mu_i^{\rm true}, \sigma_{\mu})$, where 
$\vec{d}_{\rm EM}$ are the data associated with the EM observations, and 
$\mathcal N(\mu_i|\mu_i^{\rm true}, \sigma_{\mu})$ is a normal distribution centered around 
the true magnification value $\mu_i^{\rm true}$ of each image $i$, with a 10\% standard deviation 
for $\sigma_\mu$. 
Doing so allows us to disentangle the intrinsic $\DGW$ and magnification from $D_L^{\rm eff, i}$.
For the remainder of the discussion, we assume that the intrinsic GW luminosity distance, $\DGW$, 
has been estimated through this procedure.

\section{M\lowercase{odified Propagation Theories}}

As explained above, our tests of modified theories of gravity will be based on a comparison between
the reconstructed $\DGW$ and the luminosity distance $\DEM$ obtained by electromagnetic 
means.\footnote{Here we will focus exclusively on anomalous propagation affecting the amplitude of GWs, but 
for models that lead to dispersion, the effect on the GW \emph{phasing} of BBH signals has been used to place  
very stringent constraints 
\cite{LIGOScientific:2016aoc,LIGOScientific:2016lio,LIGOScientific:2017bnn,LIGOScientific:2019fpa,LIGOScientific:2020tif,LIGOScientific:2021sio}.  
In addition, the difference between the times of arrival of GW170817 and the associated gamma ray burst has 
enabled strong constraints on differences between the speed of gravitational waves and the speed of light 
\cite{LIGOScientific:2017zic}.} 
In the specific modified gravity models we consider -- large extra dimensions, $\Xi$-paramaterization, and varying Planck mass -- there is a non-trivial relationship between these 
two quantities, which will depend on the parameter(s) related to the deviation from GR and on the cosmological
parameters. Let us briefly recall what these relationships look like for our three models.

\label{sec:mpt}
\begin{figure}
    \includegraphics[width = 0.5\textwidth]{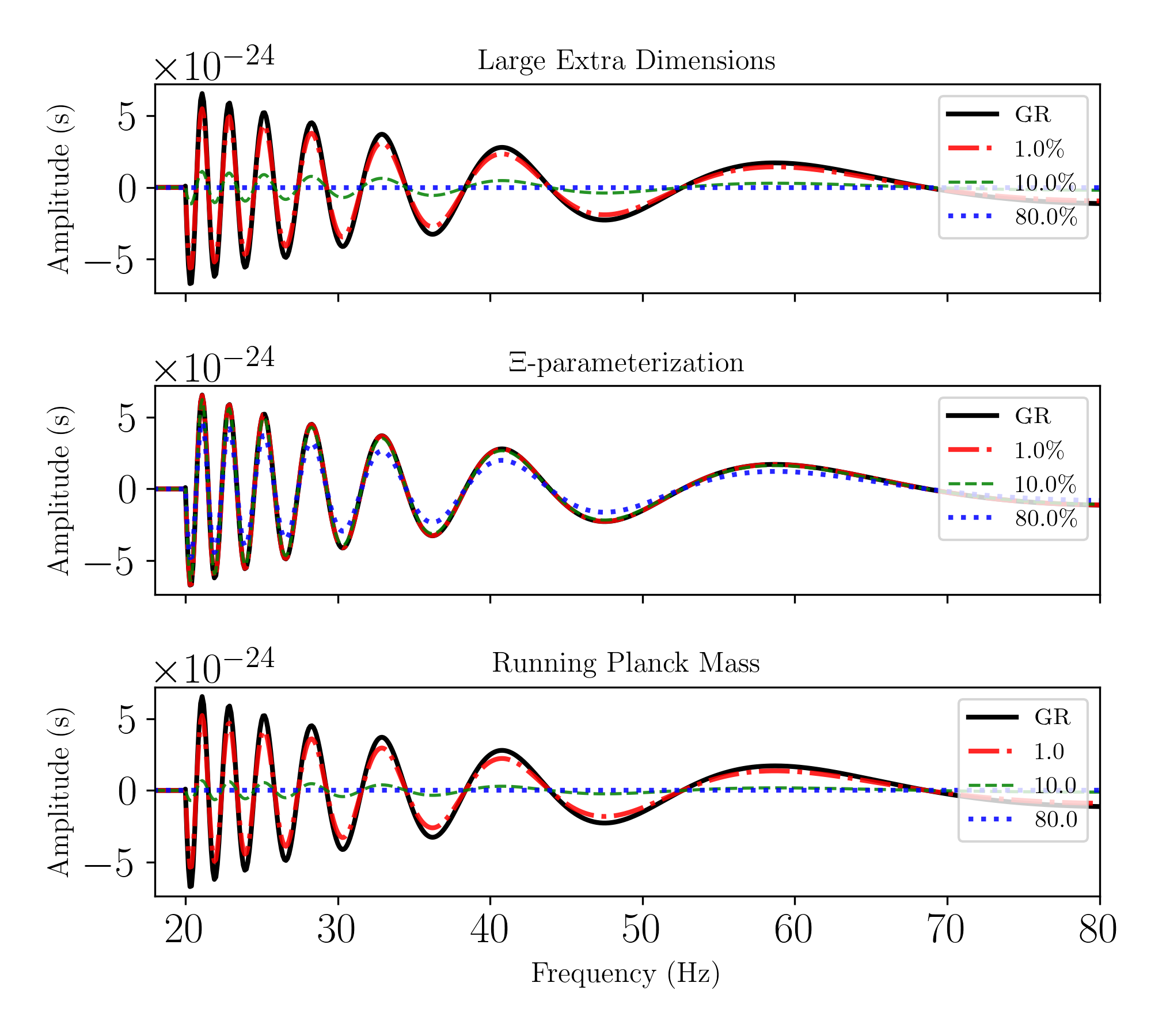}
    \caption{The effect on the frequency domain GW signal in each of the modified propagation models 
    assuming different amounts of deviations from GR denoted by different colours. 
    In these examples, the GW source is assumed to be at $\sim 5$ Gpc and the rest of the source 
    parameters are similar to those of GW150914~\cite{LIGOScientific:2016aoc}. For the running 
    Planck mass model, the deviation is absolute since one has $c_M = 0$ in GR.  
    For large extra dimensions and $\Xi$-parameterization, we consider percentage deviation in the 
    parameters $D$ and $\Xi_0$, taking the fiducial values to be $D = 4$ and $\Xi = 1$, respectively. 
    For the $\Xi$-parameterization, we arbitrarily choose $n = 1$ here, though in our subsequent analyses
    it will be a free parameter.}
    \label{fig:wf-effects}
\end{figure}

\subsection{Large extra spatial dimensions}
In theories of gravity with large extra dimensions, there is the possibility of some energy of the GWs
leaking into them \cite{Dvali:2000hr, Calcagni:2019kzo, LIGOScientific:2018dkp}, while
EM radiation is confined to the usual three spatial dimensions. This would make the detected signal 
appear weaker, leading to larger measured values for $\DGW$ than would otherwise be the case. For definiteness, we
will work with the following simple phenomenological ansatz for the relation between $\DGW$ and $\DEM$, 
based on conservation of integrated flux \cite{Deffayet:2007kf}:
\begin{equation}
	\label{eqn:D-model}
	\DGW = (\DEM(z_s, H_0))^{\frac{D-2}{2}},
\end{equation}
where $D$ is the number of spacetime dimensions and $z_s$ is the source redshift.
We will allow $D$ be a real number, with the GR value $D = 4$ as a fiducial value.  
An illustration of the effect of extra dimensions on a GW waveform is given in the top 
panel of Fig.~\ref{fig:wf-effects}.

\subsection{$\Xi-$parameterization}
Another parameterization 
was proposed in~\cite{Belgacem:2018lbp}, where the link between $\DGW$ and $\DEM$ is expressed as
\begin{equation}
	\label{eqn:xi-model}
	\DGW = \DEM(z_s, H_0)\left[\Xi_0+\frac{1-\Xi_0}{(1+z_s)^n}\right]\,.
\end{equation}
The free parameters of the model are $(\Xi_0, n)$. This parameterization is phenomenological in 
nature, but as shown in \cite{LISACosmologyWorkingGroup:2019mwx} it can be related to a large class of 
modified gravity theories, 
including Horndeski~\cite{Horndeski:1974wa} theories, 
Degenerate Higher Order Scalar-Tensor theories (DHOST)~\cite{Langlois:2018dxi}, 
and theories with nonlocally modified gravity~\cite{Maggiore:2013mea, Maggiore:2014sia,Belgacem:2017cqo}. 
When $z \ll 1$, $\DGW \simeq \DEM$. 
Therefore, similar to the extra dimension theories, we expect to observe a departure from 
GR only at large distances ($z \gtrsim 1$). For GR, 
$\Xi_0 = 1$ and $n$ is degenerate. 
In Fig.~\ref{fig:wf-effects}, middle panel, one can see an illustration of 
the effect of this modified propagation theory on the observed GW signal.

\subsection{Time-varying Planck mass}
A time-varying Planck mass is another possible cause for 
modified GW propagation. Following~\cite{Lagos:2019kds}, 
the relation between $\DGW$ and $\DEM$ can be expressed as
\begin{multline}
	\label{eqn:cm-model}
	\DGW(z) = \DEM(z_s, H_0) \times \\
	\exp{\left(\frac{c_M}{2\Omega_{\Lambda}}\ln{\frac{1+z_s}{(\Omega_{m}(1+z_s)^3 + \Omega_{\Lambda})^{1/3}}}\right)} \,,
\end{multline}
where $c_M$ is a constant that relates the rate of change of the Planck mass with the
fractional dark energy density in the Universe; for details, see \cite{Lagos:2019kds} and 
references therein. For GR, $c_M = 0$. The bottom panel of Fig.~\ref{fig:wf-effects} illustrates the change in 
a GW signal in the non-GR case.

\begin{table}
	\begin{tabular}{ c | c | c }
		\hline\hline
		Theory                 & Parameter & Priors                 \\
		\hline
		Large extra dimension        & $D$         & Uniform(3, 5)          \\
		\hline
		$\Xi$-parameterization & $\Xi_0$   & Log Uniform(0.01, 100) \\
		                       & $n$       & Uniform(0, 10)         \\
		\hline
		Running Planck mass    & $c_M$     & Uniform(-150, 150)     \\
		\hline
		
	\end{tabular}
	\caption{Deviation parameter(s) for each theory and the corresponding prior probability
	distributions used in our analyses.}
	\label{tab:priors}
\end{table}

\section{M\lowercase{ethod}}
\label{sec:methods}
In this section we provide a more detailed outline of our method to measure the parameters 
characterizing the 
deviation for each case discussed in Sec.~\ref{sec:mpt}. 

We want to measure the deviation parameters given the GW data $\vec{d}_{\mathrm{GW}}$ and 
the EM data $\vec{d}_{\mathrm{EM}}$ data associated with a strongly lensed GW 
with quadruple images whose
host galaxy has been determined. 
Let us denote the deviation parameters in all generality by $\betaMGR$. 
What we want to obtain is 
$p(\betaMGR, H_0|\vec{d}_{\mathrm{GW}}, \vec{d}_{\mathrm{EM}})$, the posterior probability distribution 
of the deviation parameters and the Hubble constant given the observed data. (As explained in the
Introduction, other cosmological parameters are given definite values.)  
Using Bayes' theorem, we can write 
\begin{eqnarray}
\label{eq:pe1}
&&p(\betaMGR, H_0 | \vec{d}_\mathrm{GW},  \vec{d}_{\mathrm{EM}}) \nonumber \\
&& = \frac{p(\betaMGR, H_0)\, p(\vec{d}_{\mathrm{GW}}, \vec{d}_{\mathrm{EM}} | \betaMGR, H_0)}{Z}
\end{eqnarray}
where $H_0$ is the Hubble constant; $p(\betaMGR, H_0)$ the prior probability distribution 
for $\betaMGR$ and $H_0$; $p(\vec{d}_{\mathrm{GW}}, \vec{d}_{\mathrm{EM}} | \betaMGR, H_0)$ the 
likelihood function; and $Z$ the evidence, whose value follows from the requirement that 
the posterior probability distribution be normalized. 
The prior distributions for $\betaMGR$ are specified in Table~\ref{tab:priors}. 
As explained in the Introduction, for $H_0$ we could in principle choose a relatively narrow
prior range based on the Planck \cite{Planck:2006aa}, SHoES \cite{Riess:2021jrx}, 
or other existing measurements \cite{Wong:2019kwg}. Instead we make the more conservative choice of
using as a prior the posterior distribution for $H_0$ obtained from the differences in 
time of arrival of the GW images, together with lens reconstruction through 
electromagnetic means. For details 
we refer to \cite{Hannuksela:2020xor,Wempe2022AObservations}; here we confine ourselves 
to recalling that what is obtained from observations is the so-called time delay 
distance $D_{\Delta t}$, which is related to $H_0$ through
\begin{equation}
\label{eqn:time-delay-dist}
D_{\Delta t}(z_l, z_s, H_0) = \frac{\int_0^{z_s} dz'/E(z')}{\int_{z_l}^{z_s} dz'/E(z')}\DEM(z_s, H_0) \, .
\end{equation} 
Here $z_l$ and $z_s$ are respectively the lens and the source redshift, and 
$E(z) \equiv \sqrt{\Omega_m (1+z)^3 + \Omega_\Lambda}$. If $D_{\Delta t}$ 
is measured, we can estimate $\DEM$ since we assume that $z_l$ and $z_s$ are known from 
the EM follow-up observations. Using the $\DEM$ measurement, $H_0$ can be estimated through
Eq.~\eqref{eqn:dl-em}. $D_{\Delta t}$ can be measured by performing lens reconstruction; 
however, owing to the computational complexity and cost, we skip the lens construction step 
and directly pick a value for the observed luminosity distance 
$D_{L, obs}^\textrm{EM}$ from a Gaussian distribution 
centred at the true value of $\DEM$ and with standard deviation $\sigma = 0.1\DEM$, allowing
us to incorporate offsets in the measurement. Next, we assume a 
Gaussian distribution around  $D_{L, obs}^\textrm{EM}$  with a 
$10\%$ standard deviation which serves as our posterior distribution for $\DEM$. 
Using the samples of this distribution together with  
Eq.~\eqref{eqn:dl-em}, we construct the prior for $H_0$. 
The 10\% standard deviation used in the previous step is motivated by the results 
of~\citet{Hannuksela:2020xor}.

To calculate the likelihood $p(\vec{d}_{\mathrm{GW}}, \vec{d}_{\mathrm{EM}} | \betaMGR, H_0)$,  
we first express it as
\begin{eqnarray}
\label{eq:likeli}
&&p(\vec{d}_{\mathrm{GW}}, \vec{d}_{\mathrm{EM}} | \betaMGR, H_0) \nonumber\\
&&= \int d\vec{\theta}\, dz_s\, p(\vec{d}_{\mathrm{GW}} | \vec{\theta})\, p(\vec{d}_{\mathrm{EM}} | z_s) \nonumber\\ 
&& \;\;\;\;\;\;\;\;\; \times \, p(\vec{\theta} | z_s, \betaMGR, H_0) \, p(z_s| \betaMGR, H_0) \,,
\end{eqnarray}
where $\vec{\theta}$ denotes the GW source parameters, $p(\vec{d}_{\mathrm{GW}} | \vec{\theta})$ 
and  $p(\vec{d}_{\mathrm{EM}} | z_s)$ are the likelihoods of 
the GW and EM data respectively, and $z_s$ is the source redshift. 
$p(\vec{\theta} | z_s, \betaMGR, H_0)$ and $p(z_s | \betaMGR, H_0)$ 
are the priors on the GW source parameters and redshift. 

Since we assume that the host galaxy has been localized, the true source redshift $z_s$ is known, and 
$p(\vec{d}_{\mathrm{EM}} | z_s)$ becomes a Dirac delta function centered on it, 
reducing Eq.~\eqref{eq:likeli} to 
\begin{eqnarray}
&& p(\vec{d}_{\mathrm{GW}}, \vec{d}_{\mathrm{EM}} | \betaMGR, H_0) \nonumber\\
&&= \int d\vec{\theta}\, p(\vec{d}_{\mathrm{GW}} | \vec{\theta}) p(\vec{\theta} | z_s, \betaMGR, H_0) \, 
p(z_s| \betaMGR, H_0) \, . \nonumber\\
&&
\end{eqnarray}

To estimate the GW likelihood $p(\vec{d}_{\mathrm{GW}}|\vec{\theta})$, we perform Bayesian parameter 
inference using nested sampling~\cite{Skilling:2006gxv}, at least for the first image. 
Subsequently we use \textsc{Golum}~\citep{Janquart:2021qov, Janquart:2023osz} 
to speed up Bayesian parameter inference for the other images. \textsc{Golum} 
can rapidly analyse lensed images by using the posterior samples of the first image as prior 
for the subsequent images, as most of the parameters for each of the four images 
are expected to be the same, apart from  
relative magnifications, rigid phase offsets, and differences in time of arrival. 

Once we have the GW likelihood, we perform the integration over the $\vec{\theta}$ for all parameters except the 
luminosity distance $\DGW$, yielding
\begin{eqnarray}
\label{eqn:step-likeli}
&&p(\vec{d}_{\mathrm{GW}}, \vec{d}_{\mathrm{EM}} | \betaMGR, H_0) \nonumber\\ 
&& = \int dD_L^{\mathrm{ GW}} \, p(\vec{d}_{\mathrm{GW}} | D_L^{\mathrm{GW}}) \,
 p(D_L^{\mathrm{GW}} | z_s, \betaMGR, H_0) \nonumber\\ 
&& \;\;\;\;\;\;\;\;\; \times \, p(z_s| \betaMGR, H_0) \,.
\end{eqnarray}

The prior $p(\DGW | z_s, \betaMGR, H_0)$ reduces to a Dirac delta function as we exactly  
know $\DGW$ given the values of $z_s$, $\betaMGR$, $H_0$ and the modified gravity model (Eqs.~\eqref{eqn:D-model}, \eqref{eqn:xi-model} and \eqref{eqn:cm-model}). 
Therefore, integrating with respect to $\DGW$ leads to
\begin{equation}
	\label{eq:redu-likeli}
	p(\vec{d}_{\mathrm{GW}}, \vec{d}_{\mathrm{EM}} | \betaMGR, H_0)  = p( \vec{d}_{\mathrm{GW}} | \DGW) p(z_s|\betaMGR, H_0).
\end{equation}
Substituting Eq.~\eqref{eq:redu-likeli} into Eq.~\eqref{eq:pe1} we can obtain the posterior distributions for $\betaMGR$ and $H_0$.

In what follows, we assume binary black hole coalescences with component mass distributions drawn from the \textsc{PowerLaw+Peak} in~\cite{KAGRA:2021duu}.
Our GW waveform model is IMRPhenomXPHM~\cite{Pratten:2020ceb}, with black hole spin magnitudes distributed uniformly 
between $0$ and $1$, and spin directions uniformly on the sphere.
The distribution of the redshifts of the BBH and the galaxy lenses (modelled as singular power law isothermal ellipsoids with external shear) is obtained from~\citet{Wierda:2021upe}.
The fiducial values of $\betaMGR$ are equal to their GR values. The fiducial value of the Hubble constant is 
$H_0 = 67.4$ km $\text{s}^{-1}$ $\text{Mpc}^{-1}$, and $\Omega_m = 0.315$. 
The lensed GWs were analyzed using \textsc{Golum}~\cite{Janquart:2021qov, Janquart:2023osz} and \textsc{Dynesty}~\cite{Speagle2019Dynesty:Evidences} to produce the $\DGW$ posteriors along with other source parameters. 
Our detector network consists of two LIGO~\cite{LIGOScientific:2014pky}, the Virgo~\cite{VIRGO:2014yos}, the KAGRA~\cite{KAGRA:2020tym}, and the LIGO-India~\cite{Unnikrishnan:2013qwa} detectors where the detection threshold on the network SNR is 8.  
Results obtained using lensed events will be compared with what can be obtained from the GW observation of 
the BNS merger GW170817 together with its host galaxy identification~\cite{LIGOScientific:2017adf}.
For GW170817, we use the $\DGW$ posterior sample from the corresponding data release~\cite{LIGOScientific:GWTC1DR}. 
For this event we cannot construct the prior on $H_0$ for GW170817 using the method which we used for lensed events; 
therefore we use Planck 2018~\cite{Planck:2018vyg} results when analyzing it. 

\begin{figure}
    \includegraphics[width = 0.5\textwidth]{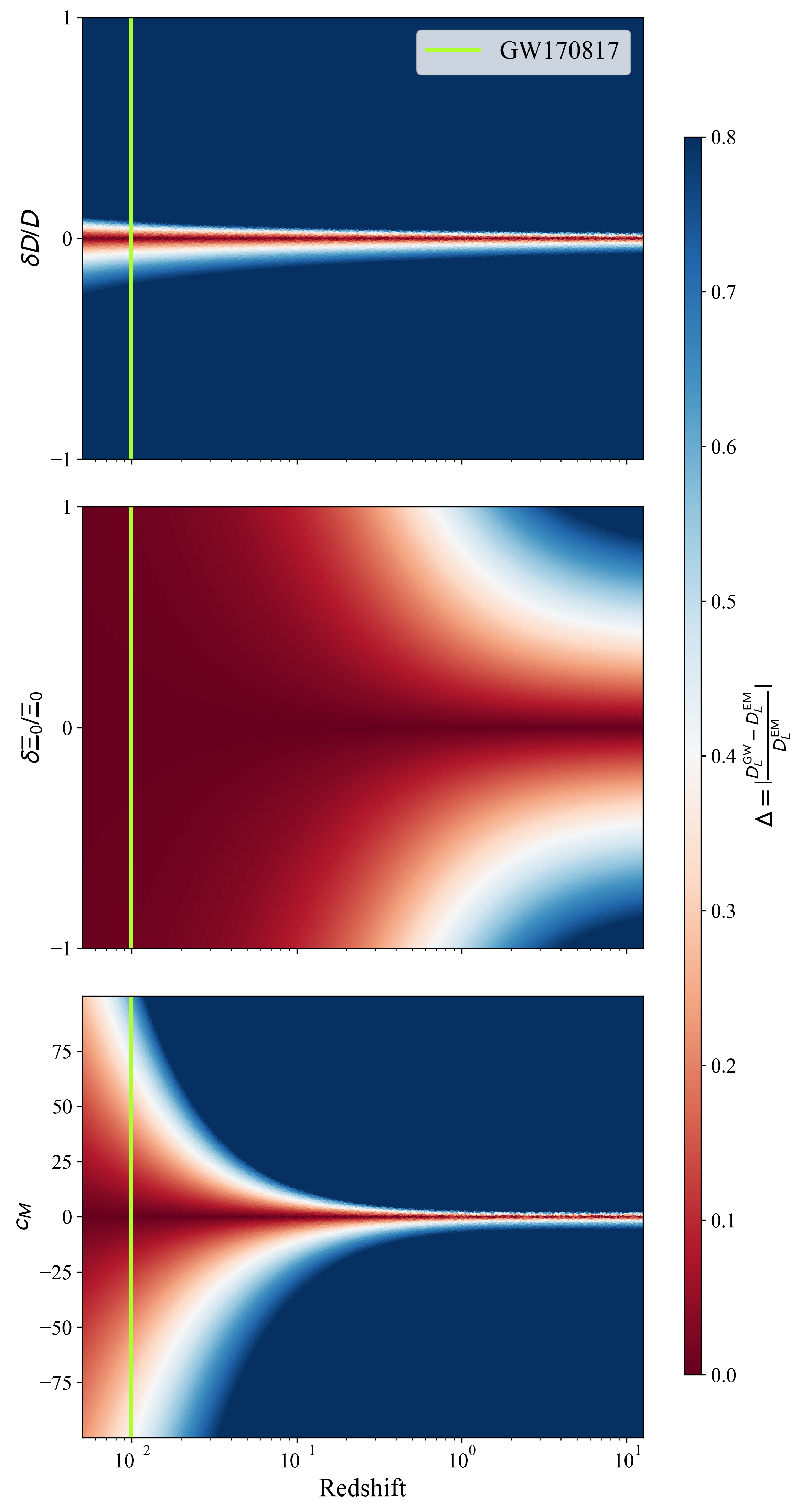}
    \caption{
    The fractional difference $\Delta \equiv |(\DGW - \DEM)/\DEM|$ (color) between $\DGW$ and $\DEM$ as a function of source redshift 
    (horizontal axis) and deviation parameter (vertical axis). The $\delta D/D$ (top panel) and $\delta\Xi_0/\Xi_0$ (middle) 
    refer
    to changes in respectively $D$ and $\Xi_0$ relative to their fiducial values $D = 4$ and $\Xi_0 = 1$
    (with $n = 1$ for the latter case), whereas for 
    $c_M$ (bottom panel) we use the value of the parameter itself. In the blue (red) regions the impact 
    of the deviation parameter on the relation between $\DGW$ and $\DEM$ is larger (smaller). 
    At the redshift of GW170817 (green vertical line), 
    for the $\Xi$-parametrization and varying 
    Planck mass, $\Delta$ is smaller than at high redshifts, already suggesting that strong lensing 
    measurements, which access the high-redshift regime, are likely to lead to better constraints on these deviation 
    parameters.  
    On the other hand, the effect of extra dimensions is less sensitive to  
    redshift, and measurements of $D$ are not expected to improve as much as for the other two cases.
     }
    \label{fig:comb-const}
\end{figure}

\section{R\lowercase{esults}}
\label{sec:results}
\begin{figure}
    \includegraphics[width = 0.45\textwidth]{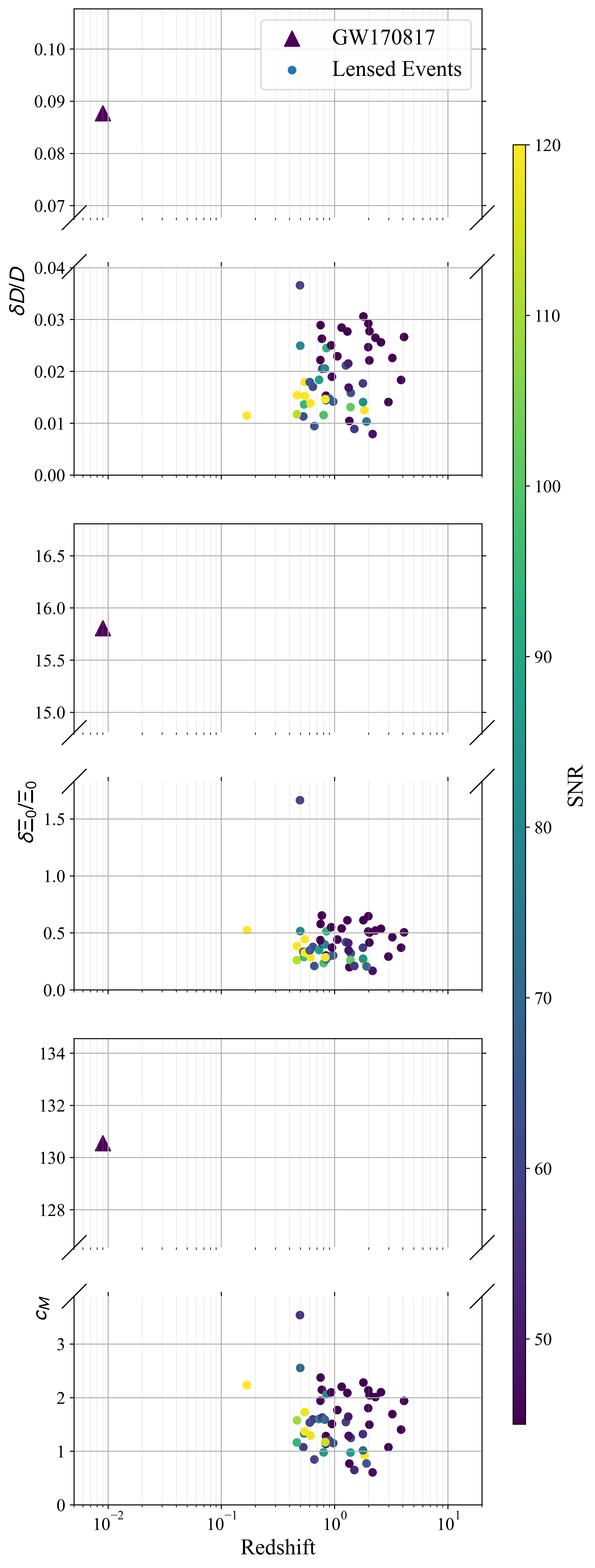}
    \caption{90\% confidence intervals for measurements of $\delta D/D$, $\delta \Xi_0/\Xi_0$ 
    (defined as in Fig.~\ref{fig:comb-const}) and $c_M$. The dots refer to results from 
    quadruply lensed events, whose source redshifts can be read off from the horizontal axis;  
    in each case the colors indicate the combined signal-to-noise ratios (SNRs) from the four images.
    The triangle indicates bounds from GW170817. Even lensed events with combined SNR similar to that of 
    GW170817 (which was $\simeq 32.4$) yield considerably better constraints on deviation parameters, 
    again underscoring the benefit of being able to access the high-redshift regime.}
\label{fig:comb-ngr-params}
\end{figure}

Before diving into the full parameter estimation results, we first look into how the relative 
difference $\Delta = |\DGW-\DEM|/\DEM$ varies as a function of $z_s$ and $\betaMGR$, to 
help us understand how large the imprint of various deviations will be. Values for $\Delta$
are indicated by the color coding in Fig.~\ref{fig:comb-const}. Here $\DEM$ is calculated 
for a range of values for redshift (horizontal axis), and $\DGW$ is computed using 
Eqs.~\eqref{eqn:D-model}-\eqref{eqn:cm-model} for a variety of (relative) deviation parameters 
(vertical axis).   
If $\Delta$ is small (red regions), there may only be a negligible imprint in 
the departure from GR even if the deviation parameter differs significantly from 
its GR value. In the blue regions, we have a better chance of observing a deviation from GR if it is 
present. 

The green vertical line shows the measured redshift of the host galaxy of GW170817 
($z \simeq 0.009783$ \cite{Levan:2017ubn,Hjorth:2017yza}). For the extra dimensions model, 
the line is mainly in the blue region, making the imprint of the deviation relatively large even 
for relatively small departures from the fiducial value of $D = 4$. However, 
for the given ranges of the $\Xi_0$ and $c_M$ parameters, GW170817 stays mostly in 
the red regions, making it more difficult to find the corresponding deviations from GR. 
In the latter two cases, higher redshifts than that of GW170817 are needed to have significantly 
better bounds on $\Xi_0$ and $c_M$, and this is what GW lensing will provide. 

In Fig.~\ref{fig:comb-ngr-params}, we present the results obtained from a 
detailed simulation, as explained in the previous section. We consider a total of 55 GW events for the analysis. Each dot in Fig.~\ref{fig:comb-ngr-params} corresponds to a simulated strongly lensed GW event with quadruple images, at a given 
source redshift (horizontal axis), analyzed as described in Sec.~\ref{sec:methods}.
The true values of deviation parameters are set equal to their GR values. 
The vertical axis indicates the 90\% confidence intervals for relative deviations in $D$ (top) 
and $\Xi_0$ (center), and 
for the absolute deviation in $c_M$ (bottom), as the latter parameter is zero in GR. 
Since in the $\Xi$-parameterization, the parameter $n$ is unconstrained when $\Xi_0$ 
equals its fiducial value of 1, we do not show results for it here, though it was treated as a free 
parameter in our measurements. Finally, the color coding shows the combined 
SNR from the four images, i.e.~the quadrature sum of the SNRs of the individual images. 
Also included are results from GW170817.

The results are in qualitative agreement with Fig.~\ref{fig:comb-const}. 
In particular, for $\Xi_0$ and $c_M$ the advantage of being able to access higher redshifts is 
clearly in evidence, with bounds improving over those of GW170817 by factors of 
up to $\mathcal{O}(10)$ and $\mathcal{O}(100)$, respectively. By contrast, the bounds on $D$ improve by up to a factor 
of $\sim 5$. The differences in improvement can be explained by  
the qualitative predictions of Fig.~\ref{fig:comb-const} where 
$\Delta$ follows a steep gradient for $\Xi_0$ (center) and $c_M$ (bottom) 
but a shallow one for $D$ (top). 

We note that for the strongly lensed events in our catalog, the combined SNR from the four images 
tends to be higher than that of GW170817, which can also improve 
the measurement accuracy on $\DGW$ and $\betaMGR$.  
Indeed, the measurement of the parameters is done using combined information 
from the different images, increasing the effective SNR used to infer the parameters values. 
However, in Fig.~\ref{fig:comb-ngr-params} 
we observe that lensed events with SNR similar to GW170817 (which has SNR $\simeq$ 32.4 \cite{LIGOScientific:2017vwq}) 
can measure the $\betaMGR$ more accurately compared to the latter as the lensed events are placed at high redshifts. 
Therefore, an increment in the distance made accessible by strong lensing is indeed the dominating factor 
in the improvement of measurement accuracies. 
	
For the $\Xi$-parameterization, bounds we obtain from our simulated lensed events are consistent with the results of 
Finke et al.~\cite{Finke:2021znb}. Let us also make a comparison with existing bounds from actual measurements. We have 
already mentioned the improvements of bounds from lensing with respect to measurements done with GW170817. 
In Mastrogiovanni et al.~\cite{Mastrogiovanni:2020mvm}, bounds were obtained for the three models considered here, 
by combining information from GW170817 and its EM counterpart with information from the BBH
signal GW190521, in the latter case assuming that a particular EM flare observed by the Zwicky 
Transient Factory (ZTF) \cite{ZTF-BBH} was associated with the 
BBH merger. Since GW190521 originated at a redshift of $\simeq 0.8$ \cite{LIGOScientific:2020iuh}, 
adding this event brings the bounds on deviation parameters closer to what we find for lensed events; 
for example, they report $\delta\Xi_0/\Xi_0 \lesssim 3 - 10$ depending on assumptions made, to be compared 
with the bounds in Fig.~\ref{fig:comb-ngr-params}.\footnote{However, it should be noted that the association
of GW190521 with the EM flare of \cite{ZTF-BBH} is by no means conclusive; see e.g.~\cite{Ashton:2020kyr}.} 
When specific alternative theories of gravity are assumed, studies based on the Cosmic Microwave Background
and large structure formation can lead to bounds on $c_M$ that are similar to the ones for lensed events; 
see e.g.~\cite{Noller:2018wyv} and the discussion in \cite{Lagos:2019kds}. Finally, methods have developed 
that exploit the observed population properties of binary black hole coalescences using gravitational
wave data only, in terms of e.g.~redshift and mass distributions 
\cite{Ezquiaga:2021ayr,Leyde:2022orh,MaganaHernandez:2021zyc}. Depending on the assumptions made, these  
can be competitive with bounds on anomalous GW propagation that we project for lensed GW events with 
host galaxy identification.

\section{C\lowercase{onclusions and Future Directions}}
\label{sec:conclusions}
Strong lensing of GWs could be detected in the near future, and there are various applications to be developed thanks 
to the additional information it can provide. Here we have focused on the fact that, under favorable circumstances, 
a quadruply lensed GW event together with EM observations can enable the identification of the host galaxy of 
a BBH event. In turn, this opens up the possibility of constraining alternative theories of 
gravity that predict anomalous GW propagation, by comparing the luminosity distance $\DEM$ that is obtained 
electromagnetically with the luminosity distance $\DGW$ obtained from the GW if the amplitude of the latter
is assumed to be proportional to $1/\DGW$. Three heuristic relationships between 
$\DGW$ and $\DEM$ were considered, motivated by large extra spatial dimensions, a variable Planck mass, 
and the so-called $\Xi$-parameterization which captures anomalous propagation effects in a variety of 
alternative theories. 

To study what kinds of constraints can be put on these non-GR models using lensed GW events, we set up 
an extensive simulation, making use of realistic lens and BBH source populations to arrive at plausible 
distributions for the properties of quadruply lensed events. We performed Bayesian inference on each of the
simulated GW events to obtain posterior density distributions for their parameters. Due to the associated 
computational complexity and cost, we did not directly perform lens reconstruction, but instead assumed Gaussian probability 
distributions for image magnification measurements used in the reconstruction of $\DGW$, 
as well as for reconstructed electromagnetic luminosity distances, with widths informed by 
current astrophysical expectations \cite{Hannuksela:2020xor,Wempe:2022zlk}. The latter aspect is something we aim to treat 
in more depth in a future study. Similarly, the relation between $\DGW$ and $\DEM$ involves cosmological parameters; 
in this work we only let $H_0$ be a free parameter, but the effect of uncertainties in the other parameters is also 
worth investigating. On the other hand, in this study we used as a prior on $H_0$ the posterior density distribution obtained 
from time delay measurements and lens reconstruction, which is typically considerably wider 
than the ranges for $H_0$ obtained from either Planck or SHoES \cite{Hannuksela:2020xor}. Because of 
the degeneracy between $H_0$ and the deviation parameters, bounds on the latter are to a large
extent set by the prior range of $H_0$ \cite{Lagos:2019kds}, which pushes our constraints on alternative 
theories towards the conservative side.\footnote{When analyzing the lensed events with a prior from Planck 2018 
\cite{Planck:2018vyg} we obtain bounds that are a factor of $\sim 2$ tighter.}

Comparing with results from GW170817 and its EM counterpart (for which we did use the much more 
narrow $H_0$ prior from Planck 2018 \cite{Planck:2018vyg}), we clearly see the effect of strongy lensed GWs 
from BBH typically originating from much higher redshifts. The latter improves the measurability of 
anomalous propagation, since it increases with distance. In the case of extra dimensions, 
modest gains by up to a factor of $\sim 5$ are seen, but for the $\Xi$-parameterization this becomes
$\mathcal{O}(10)$, and for $c_M$ as much as $\mathcal{O}(100)$. 

Previous GW-based measurements on anomalous propagation models 
\cite{LIGOScientific:2018dkp, Pardo:2018ipy, Belgacem:2018lbp, Lagos:2019kds, Mastrogiovanni:2020mvm} 
have utilized GW170817 with its EM counterpart (and GW190521 under the assumption that an EM flare
seen by ZTF was an EM counterpart to this BBH event). Until the advent of third-generation GW observatories
such as Einstein Telescope \cite{Hild:2010id,ET,Maggiore2019ScienceTelescope} and Cosmic Explorer 
\cite{CE_PSD,Reitze:2019iox}, GW signals from binary neutron star inspirals will only be seen to redshifts $z \ll 1$ 
\cite{aplus_sensitivity}, and the definitive identification of transient EM counterparts to stellar mass BBH events
may remain elusive. Other methods based on the population properties of binary black holes inferred from 
GW data alone have been shown to considerably improve over bounds from multimessenger observations of GW170817
\cite{Ezquiaga:2021ayr,Leyde:2022orh,MaganaHernandez:2021zyc}. What we have demonstrated here is that a single 
fortuitous discovery of a quadruply lensed  GW event in conjunction with EM observations of lensed galaxies may 
give access to the high-redshift regime, again enabling significantly stronger constraints on models of 
anomalous GW propagation.

\section*{A\lowercase{cknowledgments}}

H.N., J.J., K.H., and C.V.D.B. are supported by the research programme
of the Netherlands Organisation for Scientific Research (NWO).
L.H. is supported by the Swiss National Science Foundation grant 199307, as well as the European Union’s Horizon 
2020 research and innovation programme under the Marie Skłodowska-Curie grant agreement No 945298-ParisRegionFP. 
She is a Fellow of Paris Region Fellowship Programme supported by the Paris Region, and acknowledges the support 
of the COST Action CA18108. The authors are grateful for computational resources provided by the LIGO 
Laboratory and supported by the National Science Foundation Grants No.~PHY-0757058 and No.~PHY-0823459. 
This research has made use of data, software and/or web tools
obtained from the Gravitational Wave Open Science Center (https://www.gw-openscience.org), a
service of LIGO Laboratory, the LIGO Scientific Collaboration and the Virgo Collaboration.
LIGO is funded by the U.S. National Science Foundation. Virgo is funded by the French
Centre National de Recherche Scientifique (CNRS), the
Italian Istituto Nazionale della Fisica Nucleare (INFN)
and the Dutch Nikhef, with contributions by Polish and Hungarian institutes.

\bibliography{references}

\end{document}